# PUBLIC ENGAGEMENT IN ACTION: DEVELOPING AN INTRODUCTORY PROGRAMMING MODULE FOR APPRENTICES

**J. Yang, M. Seyedebrahimi, M. Low, H. Heshmati**

*University of Warwick (UNITED KINGDOM)*

## Abstract

Programming is a crucial skill in today's world and being taught worldwide at different levels. However, in the literature there is little research investigating a formal approach to embedding public engagement into programming module design. This paper explores the integration of public engagement into an introductory programming module, at the University of Warwick, UK, as part of the Digital and Technology Solutions (DTS) degree apprenticeship.

The module design follows a 'V' model, which integrates community engagement with traditional programming education, providing a holistic learning experience. The aim is to enhance learning by combining programming education with community engagement. Apprentices participate in outreach activities, teaching programming and Arduino hardware to local secondary school students. This hands-on approach aligns with Kolb's experiential learning model, improving communication skills and solidifying programming concepts through teaching. The module also includes training in safeguarding, presentation skills, and storytelling to prepare apprentices for public engagement. Pedagogical techniques in the module include live coding, group exercises, and Arduino kit usage, as well as peer education, allowing apprentices to learn from and teach each other.

Degree apprentices, who balance part-time studies with full-time employment, bring diverse knowledge and motivations. The benefit of public engagement is that it helps bridge their skills gap, fostering teamwork and creating a positive learning environment. Embedding public engagement in programming education also enhances both technical and soft skills, providing apprentices with a deeper understanding of community issues and real-world applications. Our design supports their academic and professional growth, ensuring the module's ongoing success and impact.

## 1 INTRODUCTION

Programming, also referred to as coding, is a fundamental skill in today's world. Raman Nambiar noted the value of programming in the book chapter '*Coding as an essential skill in the Twenty-First Century*' [1], with the challenge being that young people are good at consuming technology but lack skills to create the technology.

Programming is taught at various levels, ranging from beginner-friendly drag-and-drop coding styles like MIT Scratch [2], to advanced topics like deep learning and computer vision for more advanced learners. The teaching methods vary, from tailored personal tutoring systems like Khanmigo [3] to enrolment of hundreds in the same MOOC (Massive Open Online Courses) [4]. The impact of Generative AI (GAI) on coding is yet to be fully evaluated; however, new tools created using GAI, such as GitHub Copilot, have significantly simplified coding by allows developers to generate standardised code quickly and focusing on innovative creations [5].

The Royal Society Computing Snapshot Report '*Policy Briefing on teachers of computing*' in 2019 [6], recognised the challenge that schools face in this area, noting there is a decline in the number of computing teachers and fewer schools are offering computing courses. To address these challenges, schools can be receptive to activities that offer a means by which their students can gain valuable experience. Outreach and public engagement activities that involve programming are popular. These activities vary with age groups, ranging from contests like IEEEXtreme, organised by the Institute of Electrical and Electronics Engineers (IEEE) [7], to the Google Summer of Code and Microsoft Hackathon [8]. For younger children, activities often involve hardware like the BBC micro:bit [9].

The National Co-ordinating Centre for Public Engagement (NCCPE) [10], is a useful source of guidance for Higher Education organisations wishing to take a strategic approach to public engagement activities and is funded by the UK Research and Innovation (UKRI), the devolved Higher Education funding bodies, Arts Council England and Wellcome [11]. One element of NCCPE's work is to highlight the value of 'engaged teaching', which involves the development of activities that enhance students' engagement skills [12].

Several factors contribute to the popularity of programming as an outreach activity across different age groups in the educational spectrum. Resnick and Rosenbaum in *'Designing for Tinkerability'* [13] identified that physical computing activities helped the development of skills in maths, engineering and computation, as well as aiding the development of soft skills like creative thinking and collaboration which were skills children needed to actively participate in society. Concepts like branching, iterations, and exception handling have many real-world applications in various scenarios [14]. Additionally, programming lays the foundation and serves as a vehicle for delivering knowledge and skills training in subjects beyond the STEM areas. These are valued programming skills that can also be valuable for public engagement activities [15].

Few studies in the literature investigate a formal approach to embedding public engagement into programming module design. In this paper, we report on our practice of embedding community engagement into an introductory programming module design. We detail our module design process about how outreach can be used to increase classroom engagement and serve as a key motivator.

## 2  PEDAGOGICAL FRAMEWORK: PROGRAMMING EDUCATION

The teaching and learning of programming follow the general rules of human cognition [16]. One set of principles, Rosenshine's principles [17], is well-known among educators and widely applicable due to its general relevance. In fact, the ideas behind these principles are so universal that they also appear in other areas in slightly different forms. For example, the principles of daily review and task scaffolding in education are akin to daily scrums and sprint planning in the context of agile software development [18].

Specific techniques are employed in programming pedagogy. For instance, live coding, where teaching staff develop programmes in front of students, help to guide students through solving a programming problem in real-time. Brown and Wilson [19] identifies multiple benefits, allowing students to follow each step of the process, including error-fixing, leading to successful outcomes [19]. Our previous study [20] has shown that short code snippets, when delivered interactively with personalised explanations, are an effective teaching method. This approach aligns with Rosenshine's principles of using worked examples and presenting new material in small steps. Our previous research has also demonstrated the promise of advanced activities like virtual reality (VR) based education in engaging learners [21]. Yet, uncertainty persists regarding the availability of such resources for educators on a broader scale and their implications for the education system. Therefore, while activities like VR hold potential benefits, the approaches and activities outlined in this paper, which demand lower-end resources, may present a more practical and accessible alternative.

Contributing pedagogy embraces students' involvement and as explored by Hamer, Sheard, Purchase and Luxton-Reilly [22] it is a pedagogy that encourages students to contribute to the learning and to value the contributions of others. Moreover, the concept of 'peer education' [19] is notable in computer science pedagogy. Through peer-to-peer learning and discussions, students can organise their knowledge more coherently and reinforce forgotten elements, thereby reducing their 'forgetting curves' [23]. However, peer education often occurs in classroom settings like group discussions or debates, and it's uncommon outside the classroom. Believing in the adage 'while we teach, we learn,' our programming module design incorporates preparing HE apprentices to teach programming concepts to younger pupils. This iterative process of learning, preparing to teach, and ultimately teaching consolidates their knowledge and is more effective than traditional classroom-based peer education.

Different programming languages cater to different age groups and technologies. Python has been the fastest growing programming language [24] and still remains huge popularity. However, high-level general-purpose languages like Python can be conceptually challenging and abstract. Educators have attempted to mitigate this through visualisation tools such as the Python Turtle library [25]. For younger learners, visually appealing platforms like MIT Scratch [2] use a drag-and-drop interface with colourful, shaped blocks for coding. MIT App Inventor [26] offers a similar approach for Android programming, allowing interactive products on mobile devices. Sphero BOLT [27] takes this further, enabling JavaScript coding on tablets to control a ball-shaped robot. Literature highlights the valued role of C++ and hardware kits such as Arduino in STEM education and outreach activities [28]. Arduino's affordability compared to other devices, its collection of low-cost sensors and outputs, and the abundance of resources available for further learning contribute to its widespread application across various domains. Robotics is another notable context to facilitate the comprehension of computer programming [29]. Nonetheless, it may involve multiple programming languages, which presents a challenge as it is uncommon for more than one language to be taught within the same course or outreach activity.

## 3 MODULE DEVELOPMENT: INCORPORATING PUBLIC ENGAGEMENT

The programming module being discussed in this paper is called ''Smart Solutions Development I', hereafter referred to as WM164, taught at the WMG, the University of Warwick. It is part of the Digital and Technology Solutions (DTS) degree apprenticeship, offered by various employers and taught at different universities in the UK. All DTS degrees follow the standard defined by the UK Institute for Apprenticeship and Technical Education (IfATE), entails a four-year study programme for apprentices. Upon completion and successful passing of the end-point assessment, apprentices are awarded a BSc degree. At University of Warwick, WM164 is a key first-year module as it lays the groundwork in basic programming concepts, specific C++ and Python syntax, and relevant data and software skills for apprentices, enabling them to excel in their academic and professional roles.

### 3.1 Dual Roles: Students as Learners and Outreach Ambassadors

The core of our module is our students, and the primary aim of our module design is to enhance their overall experience. It is important to note that our students are not conventional undergraduate students, but degree apprentices. In conventional university settings, full-time students interact almost exclusively with the university and typically possess similar levels of knowledge and skills upon entering courses. However, degree apprentices must engage with the university, employer, and other stakeholders, including regulatory and standardisation bodies. As illustrated in Figure 1, incorporating outreach and community engagement activities will enhance these interactions.

Apprentices are part-time students and full-time employees. They are fully immersed in real-world commercial environments, and work-based learning and on-the-job training are integral to their apprenticeship. While these work experiences are valuable, degree apprentices do possess certain characteristics that are not normally seen in their full-time counterparts, which must be addressed when designing the curriculum.

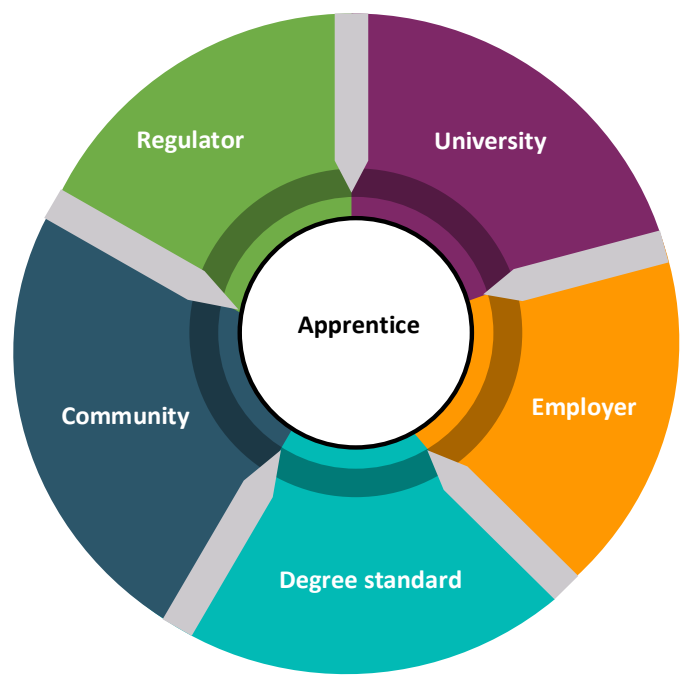

*Figure 1 The level of interactions for degree apprentices are more complicated compared to their full-time counterparts, and public engagement adds a potential further dimension of interaction.*

- First, some being mature, apprentices often have varied and sometimes divergent prior knowledge and experience, making the teaching of subjects like programming more challenging compared to full-time students.
- The motivational factors driving apprentices to strive for academic excellence differ from those of full-time students as well. They tend to have specific intent to acquire skills that directly enhance their career prospects, rather than academic achievement for its own sake [30].
- It is also felt that their community building is less effective given their part-time nature, which subsequently impact the peer learning and the forming of positive dynamics among the learners.

After learning programming and developing public engagement skills, apprentices are the ones who design and lead the engagement activities during outreach events at schools. During the classroom exercises, students learn specific programming knowledge and prepare themselves for the public engagement activities. While the knowledge and skill sets developed in technical and public engagement parts of the module are distinct, they converge at the public engagement events. By working together to design and run school outreach activities, students gain more exposure to each other and the community. This collaboration helps to minimise the skills gap, contributes to establishing long-term career goals, and fosters community building.

## 3.2 Integrating Public Engagement into Curriculum Design

Originally, the WM164 module followed a typical university structure with lectures and computer lab sessions. In these sessions, apprentices engaged with programming challenges. This approach blended abstract conceptualisation with experimentation and concrete experience, aligning with Kolb's experiential learning model [31]. However, the shift from traditional lecturing, where the entire class listens to the tutor in front of a large screen, to individual-based computer labs where each student works individually on a computer, has resulted in a significant reduction in the physical interactions among the learners that are typically common in STEM labs. This shift resulted in a focus on technical aspects rather than communication, limiting students' opportunities to articulate their work clearly, especially for relatively complex concepts such as iterations. In summer 2023, the WM164 module underwent a redesign. Overall, the redesign aimed to enhance the apprentices' learning experience by integrating public engagement-related skill development into the programming curriculum. It was decided to use programming air quality monitoring kits as the topic for community engagement, which has the following benefits: One is that the secondary school pupils would have some familiarities with the topic, they would have learned fundamental coding concepts and have heard about different discussions about climate change from different channels. In addition, programming and the wider software world can easily trigger interests and conversations with the younger generation.

The module's redesign also considered programming pedagogy, like benefiting from physical interactions in a computer laboratory and the effectiveness of peer learning in group exercises [32]. The redesigned module's structure followed a 'V' model, connecting learning and teaching perspectives, as shown in Figure 2. The process began with embedding community engagement learning as one of the module's learning outcomes (LOs), alongside existing programming-related LOs such as language syntax and grammar. LOs were aligned with the apprenticeship and AHEP (Accreditation of Higher Education Programmes) standards defined by the IfATE. The LOs were delivered through lectures, including live coding, laboratory, and group activities. Unique elements of the redesign included:

- A dedicated lecture focusing on essential safeguarding considerations for school engagement, along with practical tips for delivering captivating sessions.
- Lab sessions utilising Arduino kits to visualise programming concepts.
- Group exercises designed for technical problem-solving and practising transferable skills like presentation.
- Community engagement visits to local secondary schools, where apprentices taught programming and Arduino hardware for air quality data collection and analysis.

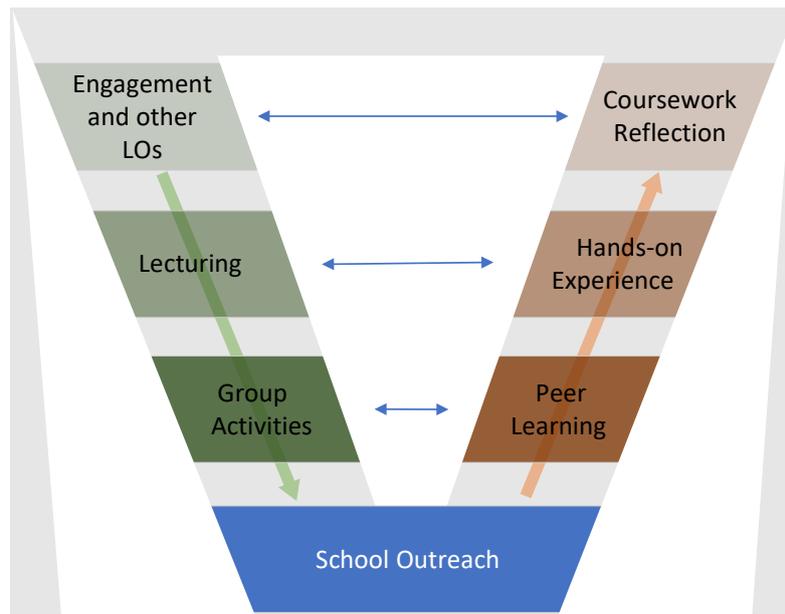

*Figure 2 The module redesign incorporates public engagement and builds on top of existing pedagogy. Teaching (green parts) and learning (brown parts) intersects at school outreach activities where apprentices need to deliver public sessions.*

During the module, apprentices also received training on transferable public engagement and school outreach skills in three different aspects:

1. Safeguarding considerations, for example, the appropriateness of language and behaviour.
2. Slide design, focusing on simplicity and engaging.
3. Storytelling, including things such as knowing the audience and avoiding jargons. From a learning perspective, apprentices experience the module as an integrated whole, comprising programming and public engagement elements. The LOs are assessed through reflection in coursework reports, which completes the 'V' model.

## 4  DISCUSSIONS

### 4.1  Considerations for Scalability and Sustainability

At present, there are at least two key considerations that need to be addressed to ensure the sustainable future running of the module: facilitating the participation of all apprentices in the class and evaluating the project's success. Both of these aspects must be tackled before the module can continue to evolve effectively.

Firstly, although public engagement is an LO, practical constraints limit the feasibility of a large number of apprentices visiting local schools. Our initial approach was based on a first-come, first-served basis. However, this led to issues as many apprentices were enthusiastic about participating in the outreach, but the module could not accommodate everyone due to cost implications and staff availability. To streamline the coordination and support of apprentice involvement in outreach activities, a centralised registration system could be implemented. This system will allow a surplus of interested apprentices to register. Following which, a peer-review process will be conducted to select apprentices to participate in the outreach activity. This approach will not only ensure efficient coordination but also contribute to a better integration of the activity within the LOs of the module.

Secondly, evaluating the success of outreach projects requires clear criteria that go beyond simply counting engaged students. While the number of participants is a significant metric, embedding outreach within the teaching module offers a broader impact. This approach not only benefits the target audience but also provides valuable training for our apprentices, particularly when promoting the project itself becomes part of their learning experience. This expanded impact should be considered when evaluating similar projects in the future. Additionally, as the programme expands and relies more heavily on data collection through surveys, ethical considerations become paramount. Implementing safeguards, such as background checks and obtaining ethical approval before gathering information from young people, is crucial. These approvals should clearly outline the project objectives and the purpose of data collection.

## 4.2 Apprentices bridging University and Corporate Outreach

Large companies typically run their own non-profit outreach activities. One example is the Jaguar Land Rover (JLR) Schools Partnership Programme [33], where automotive engineers support local schools by running STEM-related projects and promoting the automotive industry. Degree apprentices participating in programming module school outreach activities can simultaneously support their companies' outreach missions, creating opportunities for them to act as a bridge between academics and real-world commercial opportunities.

Currently, the school outreach activities designed by our students are not industry-specific but focus more on generic environmental issues such as air quality monitoring. The advantage of this approach is that the knowledge and skills developed by our students, and delivered to local schools, are universally applicable to a wide range of problems. However, this also means that the solutions provided are often theoretical and lack maturity and efficiency. By utilising degree apprentices' unique positions, we can leverage their dual role to employ commercial-level solutions, thereby enhancing efficiency and realism in addressing real-world problems.

For example, in the context of air quality monitoring, if appropriate devices are provided, we could utilise in-car air quality monitoring systems to perform similar tasks. From the companies' perspective, collaborating with universities to design outreach activities that benefit their degree apprentices is advantageous. This not only supports the development of their own apprentices but also grants them access to interdisciplinary expertise typically available at universities.

## 5   CONCLUSIONS

In this paper, we have detailed our approach to integrating public engagement within a programming module specifically designed for a unique group of students – degree apprentices. We discussed the module's design, which not only considers the pedagogical aspects typical of a programming course but also sets public engagement as one of its key LOs. We explored the roles of our students, who not only learn and develop programming knowledge but also acquire transferable public engagement skills, enabling them to lead outreach activities at local secondary schools and raise awareness of future careers for school pupils.

It can be concluded that incorporating public engagement activities into the curriculum helps bridge the programming skill gaps of individual apprentices, as these activities tend to be more oriented towards soft skills like communication, rather than purely technical skills. This approach fosters improved team building and cultivates a more positive learning atmosphere. Additionally, public engagement provides a vital 'application field' – giving students a purpose for their studies beyond their immediate business context. This offers an additional dimension that helps apprentices gain a deeper understanding of our community, acting as an essential buffer that complements their work and study life. We also present considerations for enhancing the impact and ensuring sustainable future development of the module.

## ACKNOWLEDGEMENTS

The module redesign and school outreach activities were supported by the Warwick Institute of Engagement (WIE) through Public and Community Engagement Module Development Fund.